\documentclass[12pt]{article}
\usepackage{amsmath,amssymb,amsthm,amsxtra,overpic,bbm,bm,epsfig,subfigure}
\usepackage{hyperref}
\usepackage{mathrsfs}
\usepackage{graphicx}
\usepackage{color}
\usepackage{comment}
\usepackage{epstopdf}
\usepackage{float}
\usepackage{cite}
\usepackage{bbold}
\numberwithin{equation}{section}
\usepackage{cite}
\usepackage{hyperref}
\usepackage{slashed,stmaryrd}

\def\thefootnote{\fnsymbol{footnote}}

\begin{document}

\vspace{0.2cm}

\begin{center}
{\Large\bf Transition Probabilities in the Two-Level Quantum System with PT-Symmetric Non-Hermitian Hamiltonians}
\end{center}

\vspace{0.2cm}

\begin{center}
{\bf Tommy Ohlsson}~$^{a,~b,~c}$~\footnote{E-mail: tohlsson@kth.se},
\quad
{\bf Shun Zhou}~$^{d,~e}$~\footnote{E-mail: zhoush@ihep.ac.cn}
\\
\vspace{0.2cm}
{\small $^a$Department of Physics, School of Engineering Sciences, KTH Royal Institute of Technology, \\
AlbaNova University Center, Roslagstullsbacken 21, SE-106 91 Stockholm, Sweden \\
$^b$The Oskar Klein Centre for Cosmoparticle Physics, AlbaNova University Center, \\
Roslagstullsbacken 21, SE-106 91 Stockholm, Sweden \\
$^c$University of Iceland, Science Institute, Dunhaga 3, IS-107  Reykjavik,  Iceland\\
$^d$Institute of High Energy Physics, Chinese Academy of Sciences, Beijing 100049, China\\
$^e$School of Physical Sciences, University of Chinese Academy of Sciences, Beijing 100049, China}
\end{center}

\vspace{1.5cm}

\begin{abstract}
We investigate how to define in a consistent way the probabilities of the transitions between the ``flavor" states of the two-level quantum system, which is described by a non-Hermitian but parity and time-reversal (PT) symmetric Hamiltonian. Explicit calculations are carried out to demonstrate the conservation of probability if a proper definition of the final state is adopted. Finally, this formalism is applied to two-flavor neutrino oscillations $\nu^{}_\mu \to \nu^{}_\mu$ and $\nu^{}_\mu \to \nu^{}_\tau$ in vacuum, where the exact PT symmetry requires the vacuum mixing angle to be maximal, which is compatible with current neutrino oscillation experiments. A possible generalization to the three-flavor case is briefly discussed.
\end{abstract}


\def\thefootnote{\arabic{footnote}}
\setcounter{footnote}{0}

\newpage

\section{Introduction}\label{sec:1}

In ordinary quantum mechanics, the Hamiltonian of a quantum system is usually assumed to be Hermitian such that its eigenvalues are real, which seems to be inevitable in the sense that the measured energies should be real numbers. In 1998, Bender and his collaborators discovered that a non-Hermitian Hamiltonian with parity and time-reversal (PT) symmetry could also possess a real spectrum~\cite{Bender:1998ke, Bender:1998gh, Bender:2002vv}, which then raises the interesting questions if a PT-symmetric non-Hermitian Hamiltonian is physically meaningful and if a PT symmetry serves as an alternative to the Hermiticity in quantum mechanics~\cite{Bender:2003gu, Bender:2005tb, Bender:2007nj}. Since then, developments of PT-symmetric non-Hermitian theories have been flourishing in two different aspects. First, great efforts have been made  in a series of papers by Mostafazadeh~\cite{Mostafazadeh:2001jk, Mostafazadeh:2001nr, Mostafazadeh:2002id, Mostafazadeh:2004mx, Mostafazadeh:2008pw} to explore the necessary and sufficient conditions for non-Hermitian Hamiltonians to have real spectra. It has been observed that a general class of pseudo-Hermitian Hamiltonians ${\cal H}$ contain real eigenvalues if there exists a linear positive-definite operator $\eta^{}_+$ in the Hilbert space such that $\eta^{}_+ {\cal H} \eta^{-1}_+ = {\cal H}^\dagger$ holds~\cite{Mostafazadeh:2008pw, Mannheim:2009zj}. Second, non-Hermitian Hamiltonians with PT symmetry have found extensive intriguing applications in optics~\cite{optics}, electronics~\cite{electronics}, microwaves~\cite{microwave}, mechanics~\cite{mechanics}, acoustics~\cite{acoustics}, atomic physics~\cite{Baker:1983zz,atomic, atomic2}, and single-spin systems~\cite{single-spin-system}. A recent review on this topic and especially the physical applications of PT-symmetric non-Hermitian Hamiltonians that have recently emerged in quantum physics can be found in Ref.~\cite{review} that discusses such applications in depth, which is, however, not the intension of this work.

In the areas of quantum field theories and elementary particle physics, there are a number of interesting examples as well. First of all, it is worthwhile to mention that the Lee model~\cite{Lee:1954iq} has been reexamined in the formalism of PT-symmetric non-Hermitian Hamiltonians and demonstrated to respect unitarity~\cite{Bender:2004sv, Jones:2007pq, Shi:2009pc}. In addition, some other fundamentally important problems, including neutrino mass generation~\cite{Alexandre:2015kra}, neutrino oscillations~\cite{JonesSmith:2009wy, Ohlsson:2015xsa}, light neutrino masses in Yukawa theory~\cite{Alexandre:2017fpq}, spontaneous symmetry breaking and the Goldstone theorem~\cite{Alexandre:2018uol}, and the Brout--Englert--Higgs mechanism~\cite{Mannheim:2018dur, Alexandre:2018xyy} have been studied in the framework of non-Hermitian theories with PT symmetry. It is worth mentioning that the formalism introduced in Refs.~\cite{Alexandre:2018uol, Mannheim:2018dur, Alexandre:2018xyy} may violate causality as first pointed out in the Lee--Wick model in Ref.~\cite{Nakanishi:1972wx}. Stimulated by such a tremendous progress in this research area, we revisit the two-level quantum system with a general PT-symmetric non-Hermitian Hamiltonian, which is completely solvable and thus one of the best examples to describe the main features of this non-Hermitian system. Although this simple system has been discussed extensively in the literature (see, e.g., Ref.~\cite{Bender:2007nj}), we have not yet noticed any explicit calculations of the probabilities for the transitions between any two quantum states. To fill this gap in the literature, we take up this assignment and apply the formalism to two-flavor neutrino oscillations in vacuum, e.g., $|\nu^{}_\mu\rangle \to |\nu^{}_\mu\rangle$ and $|\nu^{}_\mu\rangle \to |\nu^{}_\tau\rangle$. The exact PT symmetry of the Hamiltonian for atmospheric neutrino oscillations in vacuum requires a maximal mixing angle $\theta^{}_{23} = \pi/4$~\cite{Ohlsson:2015xsa}, which is perfectly allowed by all neutrino oscillation experiments. In this scenario, we demonstrate how to define the transition amplitudes and probabilities in a consistent way and show explicitly the conservation of probability. The essential idea is to introduce the ``flavor" eigenstates $|\widetilde{\nu}^{}_\alpha\rangle$ (for $\alpha = \mu, \tau$) as the ${\cal CPT}$ eigenstates, i.e., ${\cal CPT}|\widetilde{\nu}^{}_\alpha\rangle = |\widetilde{\nu}^{}_\alpha\rangle$, where ${\cal CPT}$ stands for the operator of combined charge-conjugate, parity, and time-reversal transformations.

It is worth stressing that the ``flavor" eigenstates, in which the explicit matrix form of the non-Hermitian Hamiltonian is written down, and their time evolution have never been clearly discussed before. The main focus of this work is to calculate the transition amplitudes and probabilities for these ``flavor" eigenstates, which has been motivated by the flavor oscillations of neutrinos. We admit that the introduction of the ${\cal CPT}$ eigenstates is only for practical purposes, i.e., due to maintaining conservation of probability, but it raises the question how to properly define the outgoing ``flavor" eigenstates in a consistent way in the presence of a non-unitary flavor mixing matrix in a non-Hermitian Hamiltonian system. It is certainly interesting and necessary to study this question in depth in future works.

The remaining part of this work is organized as follows. In Sec.~\ref{sec:2}, PT-symmetric non-Hermitian Hamiltonians are introduced and the concrete example of a two-level quantum system is presented to illustrate the formulation of non-Hermitian quantum mechanics. Next, in Sec.~\ref{sec:3}, two-flavor neutrino oscillations in vacuum are studied, where the oscillation amplitudes and probabilities are calculated, and possible generalizations to the three-flavor case and neutrino oscillations in matter are briefly mentioned. Note that neutrino oscillations in matter effectively fall into the class of time-dependent systems and assuming a PT-symmetric non-Hermitian Hamiltonian for neutrino oscillations in matter that system would be an example of a time-dependent non-Hermitian system. Then, in Sec.~\ref{sec:4}, we summarize our main results and conclude. Finally, in Appendix~\ref{sec:AppA}, the basic definitions of state vectors, inner products, and operators in conventional quantum mechanics based on Hermitian Hamiltonians are reviewed, while in Appendix~\ref{sec:AppB}, the general features of non-Hermitian quantum mechanics with pseudo-Hermitian or PT-symmetric Hamiltonians are presented.\footnote{It should be mentioned that the material presented in Appendixes~\ref{sec:AppA} and \ref{sec:AppB} is not novel and could be found in the references given in Appendixes~\ref{sec:AppA} and \ref{sec:AppB} or in textbooks. However, for making the presentation of this work self-consistent, we have included this material for convenience. Appendixes~\ref{sec:AppA} and \ref{sec:AppB} might be redundant to a reader familiar with both Hermitian and non-Hermitian Hamiltonians for quantum systems as well as neutrino oscillation physics.}

\section{Non-Hermiticity and PT Symmetry}\label{sec:2}

In Appendixes~\ref{sec:AppA} and \ref{sec:AppB}, we have considered the quantum mechanics with both Hermitian and non-Hermitian Hamiltonians for a quantum system with the $N$-dimensional Hilbert space and attempted to make the results as general as possible. (We will not consider infinite Hilbert spaces in this work.) To be specific, we will focus on explicit calculations in the simplest two-level quantum system (i.e., $N = 2$), for which the Hamiltonian $H$ is constant in time and space. Two linearly-independent state vectors $\{|u^{}_a\rangle, |u^{}_b\rangle\}$ are chosen to be $|u^{}_a\rangle = (1, 0)^{\rm T}$ and $|u^{}_b\rangle = (0, 1)^{\rm T}$ with ``T" denoting matrix transpose. Note that the ``flavor" state vectors $\{|u^{}_a\rangle, |u^{}_b\rangle\}$ form a complete and orthonormal basis of the two-level quantum system in the sense of $\langle u^{}_\alpha |u^{}_\beta\rangle = \delta^{}_{\alpha \beta}$ (for $\alpha, \beta$ = $a, b$), according to the Euclidean inner product. Therefore, any state vector $|\psi\rangle$ can be written as $|\psi\rangle = c^{}_a |u^{}_a\rangle + c^{}_b |u^{}_b\rangle = (c^{}_a, c^{}_b)^{\rm T}$ with $c^{}_a$ and $c^{}_b$ being complex numbers, and likewise for $|\phi\rangle = (c^\prime_a, c^\prime_b)^{\rm T}$. It is then straightforward to observe $\langle \psi|\phi\rangle = (c^*_a, c^*_b) \cdot (c^\prime_a, c^\prime_b)^{\rm T} = c^*_a c^\prime_a + c^*_b c^\prime_b$. In this ``flavor" basis, the non-Hermitian Hamiltonian ${\cal H}$ is represented by a $2\times 2$ matrix $H = (H_{\alpha\beta})$ with $H^{}_{\alpha \beta} \equiv \langle u^{}_\alpha| {\cal H} |u^{}_\beta\rangle$ (for $\alpha, \beta$ = $a, b$) being the matrix elements.

\subsection{Symmetries}

Indeed, as previously stated, in order to make sense of non-Hermitian Hamiltonians, we must restrict ourselves to a subset of non-Hermitian Hamiltonians with some symmetries. As argued in~Ref.~\cite{Bender:2007nj}, one may impose PT symmetry, which is a particular combination of the discrete space-time symmetries that have been extensively discussed in relativistic quantum theories. For definiteness, let us consider the parity operator ${\cal P}$ that is represented by the symmetric $2 \times 2$ matrix
\begin{equation}\label{eq:Pmatrx}
P = \left(\begin{matrix} 0 & 1 \cr 1 & 0\end{matrix}\right)
\end{equation}
without loss of generality \cite{Bender:2002vv}, which is uniquely defined up to unitary transformations. Note that $\det P = -1$. Furthermore, the time-reversal operator ${\cal T}$ is identified as the ordinary complex conjugation, i.e., ${\cal T} {\cal O} {\cal T}^{-1} = {\cal O}^*$ with ${\cal O}$ being an arbitrary operator. With this definition of ${\cal T}$, it naturally holds that ${\cal T}^2 = 1$. If the Hamiltonian ${\cal H}$ is required to be invariant under the ${\cal PT}$ transformation, namely, $({\cal PT}) {\cal H} ({\cal PT})^{-1} = {\cal H}$ or $\left[{\cal PT}, {\cal H}\right] = {\bf 0}$, then one can obtain the most general form for the representation matrix of the PT-symmetric non-Hermitian Hamiltonian as
\begin{equation}\label{eq:generalH}
H = \left(\begin{matrix} \rho e^{{\rm i}\varphi} & \sigma e^{{\rm i}\phi} \cr \sigma e^{-{\rm i}\phi} & \rho e^{-{\rm i}\varphi}\end{matrix}\right) \; .
\end{equation}
where $\rho$, $\sigma$, $\varphi$, and $\phi$ are real parameters and $\det H = \rho^2-\sigma^2$. Note that $H$ in Eq.~(\ref{eq:generalH}) is not symmetric, despite being PT-symmetric, and the Hermitian limit corresponds to $\varphi = n\pi$ with $n$ being an integer. Different examples of $H$ in Eq.~(\ref{eq:generalH}) have been extensively discussed in Ref.~\cite{Das:2009it}, and using the results of Ref.~\cite{Bagarello:2014}, it is possible to rewrite $H$ in terms of pseudo-fermion operators, which means that the condition of PT symmetry can be relaxed. To guarantee orthogonality of the eigenstate vectors (with respect to the ${\cal PT}$ or ${\cal CPT}$ inner products to be explained and discussed later~\cite{Bender:2007nj}), one can further demand $H$ to be symmetric, i.e.,
\begin{equation} \label{eq:Ham}
H = \left(\begin{matrix} \rho e^{{\rm i}\varphi} & \sigma \cr \sigma & \rho e^{-{\rm i}\varphi} \end{matrix}\right) \; ,
\end{equation}
where $\phi = 0$ has been enforced.\footnote{The general condition for $H$ in Eq.~(\ref{eq:generalH}) to be symmetric is $\phi = n \pi$, where $n \in {\mathbb Z}$.} Although it is possible to make the forms of parity and time-reversal operators more general~\cite{Bender:2007nj}, we persist in assuming these compact forms in order to simplify all results and discussions. As pointed out in Refs.~\cite{Mostafazadeh:2001jk, Mostafazadeh:2001nr, Mostafazadeh:2002id}, the PT symmetry is not a necessary condition for a real spectrum of a non-Hermitian Hamiltonian. The intrinsic connection between the pseudo-Hermitian Hamiltonians and the PT-symmetric non-Hermitian Hamiltonians has been clearly explained in Refs.~\cite{Mostafazadeh:2008pw, Mannheim:2009zj}.

\subsection{Eigenvalues and Eigenvectors}

The eigenvalues of the non-Hermitian Hamiltonian $H$ in Eq.~(\ref{eq:Ham}) can be immediately found by solving its characteristic equation
\begin{equation}\label{eq:eigenvalues}
E^{}_\pm = \rho \cos\varphi \pm \sqrt{\sigma^2 - \rho^2 \sin^2\varphi} \; ,
\end{equation}
and the corresponding eigenvectors are chosen as \cite{Kleefeld:2009vd}
\begin{equation}\label{eq:upm}
|u^{}_\pm\rangle = \frac{1}{\sqrt{2\cos\alpha}} \left(\begin{matrix} e^{\pm{\rm i}\alpha/2} \cr \pm e^{\mp {\rm i}\alpha/2}\end{matrix}\right) \; ,
\end{equation}
where $\sin \alpha \equiv \rho\sin\varphi/\sigma$ has been defined.\footnote{In general, there is a choice of the overall phases for the eigenvectors, namely, $|u^{}_\pm\rangle = N^{}_\pm \left(e^{{\rm i}\alpha/2}, \pm e^{\mp{\rm i}\alpha/2}\right)^{\rm T}/\sqrt{2\cos\alpha}$, where $N^{}_\pm$ are normalization factors. In particular, our choice in Eq.~(\ref{eq:upm}) follows the convention of Ref.~\cite{Kleefeld:2009vd}. Such a convention is particularly chosen to be advantageous for the ${\cal PT}$ or ${\cal CPT}$ inner products, which will be defined later, and it should be noticed that different normalization factors will appear for another choice of convention.} We have restricted ourselves to the unbroken phase of PT symmetry, namely, $\sigma^2 \geq \rho^2 \sin^2 \varphi$, and obtained two real eigenvalues $E^{}_\pm$. If this condition is not satisfied, the PT symmetry is broken and the two eigenvalues become a complex-conjugate pair~\cite{Bender:2007nj}. In addition, the vectors in Eq.~(\ref{eq:upm}) are no longer eigenvectors of $H$ in the broken phase. At the exceptional point of $\sigma^2 = \rho^2 \sin^2\varphi$, we have two degenerate eigenvalues $E^{}_+ = E^{}_-= \rho\cos\varphi$ and only one nonzero eigenvector emerges, and therefore, the eigensystem turns out to be incomplete.

Note that the eigenvectors in Eq.~(\ref{eq:upm}) have been normalized using the so-called ${\cal PT}$ inner product, see Eq.~(\ref{eq:PTnorm}). Now, $H$ can be diagonalized by the similarity transformation $A H A^{-1} = \widehat{H} \equiv {\rm diag}(E^{}_+, E^{}_-)$ with a matrix $A$, whose inverse $A^{-1}$ can be constructed by identifying the eigenvectors $|u^{}_+\rangle$ and $|u^{}_-\rangle$ as its first and second columns, respectively, i.e.,
\begin{equation}\label{eq:A}
A^{-1} = \frac{1}{\sqrt{2\cos\alpha}} \left(\begin{matrix} e^{{\rm i}\alpha/2} & e^{-{\rm i}\alpha/2} \cr e^{-{\rm i}\alpha/2} & - e^{{\rm i}\alpha/2} \end{matrix}\right) \; .
\end{equation}
Using the choice of the overall phases for the eigenvectors in Eq.~(\ref{eq:upm}), it holds that $A^{-1} = A$ and $\det A = -1$. (However, in general, note that $A^{-1} \neq A$.) Thus, it is straightforward to verify that $A^2 = \mathbb{1}_2$. The ordinary norms of the ``mass" eigenstate vectors $|u^{}_+\rangle$ and $|u^{}_-\rangle$ can be easily computed as
\begin{equation}\label{eq:norm}
\left(\begin{matrix} \langle u^{}_+| u^{}_+ \rangle & \langle u^{}_+|u^{}_-\rangle \cr \langle u^{}_-|u^{}_+\rangle & \langle u^{}_- | u^{}_-\rangle \end{matrix}\right) = \left(\begin{matrix} \sec \alpha & -{\rm i}\tan\alpha \cr {\rm i}\tan\alpha & \sec\alpha \end{matrix}\right) \; ,
\end{equation}
implying that $\{|u^{}_+\rangle, |u^{}_-\rangle\}$ does not form an orthonormal basis of the Hamiltonian system with the Euclidean inner product. On the other hand, using Eqs.~(\ref{eq:etapm}) and (\ref{eq:upm}), the completeness relation turns out to be
\begin{equation}\label{eq:complete}
\sum_{s=\pm} |u^{}_s\rangle \langle u^{}_s| = |u^{}_+\rangle \langle u^{}_+| + |u^{}_-\rangle \langle u^{}_-| = \left(\begin{matrix} \sec \alpha & {\rm i}\tan\alpha \cr -{\rm i}\tan\alpha & \sec\alpha \end{matrix}\right) = \eta^{-1}_+\; ,
\end{equation}
which leads to the metric operator
\begin{equation}\label{eq:eta+}
\eta_+ = \left(\begin{matrix} \sec \alpha & -{\rm i}\tan\alpha \cr {\rm i}\tan\alpha & \sec\alpha \end{matrix}\right)
\end{equation}
with $\det \eta_+ = 1$. The last step in Eq.~(\ref{eq:complete}) can be understood by using the second identity in Eq.~(\ref{eq:etapm}), where the basis of the eigenstate vectors $\{|\psi^{}_i\rangle\}$ should be identified with $\{|u^{}_+\rangle, |u^{}_-\rangle\}$. It is also possible to construct the ``flavor" eigenstate vectors using the ``mass" eigenstate vectors according to the similarity transformation $A H A^{-1} = \widehat{H}$, viz.,
\begin{eqnarray}
|u^{}_{a}\rangle &=& \left(A^{-1}\right)^{}_{a+} |u^{}_+\rangle +\left(A^{-1}\right)^{}_{a-} |u^{}_-\rangle = \left(\begin{matrix} 1 \cr 0 \end{matrix}\right) \; , \label{eq:u_a}\\
|u^{}_{b}\rangle &=& \left(A^{-1}\right)^{}_{b+} |u^{}_+\rangle +\left(A^{-1}\right)^{}_{b-} |u^{}_-\rangle = \left(\begin{matrix} 0 \cr 1\end{matrix}\right) \; , \label{eq:u_b}
\end{eqnarray}
which are consistent with our previous definition of $\{|u^{}_a\rangle, |u^{}_b \rangle\}$. Note that the matrix elements $\left(A^{-1}\right)^{}_{a+}$, $\left(A^{-1}\right)^{}_{a-}$, $\left(A^{-1}\right)^{}_{b+}$, and $\left(A^{-1}\right)^{}_{b-}$ in Eqs.~(\ref{eq:u_a}) and (\ref{eq:u_b}) have been labelled by two subscripts, where the first refers to the ``flavor" eigenstates $\{|u^{}_a\rangle, |u^{}_b \rangle\}$ and the second to the ``mass" eigenstates $\{|u^{}_+\rangle, |u^{}_-\rangle\}$. Although we have normalized the ``mass" eigenstates as in Eq.~(\ref{eq:upm}), the overall phases of $|u^{}_+\rangle$ and $|u^{}_-\rangle$ can be arbitrary. However, these arbitrary phases will not affect the completeness relation in Eq.~(\ref{eq:complete}) and the metric operator $\eta_+$ in Eq.~(\ref{eq:eta+}). Using the parity operator $P$ in Eq.~(\ref{eq:Pmatrx}) and the Hamiltonian $H$ in Eq.~(\ref{eq:Ham}), one can verify that $P H P^{-1} = H^\dagger$ and $\eta^{}_+ H \eta^{-1}_+ = H^\dagger$, and thus, the identity~\cite{Mostafazadeh:2002id}
\begin{equation}\label{eq:Coperator}
P^{-1} \left(\eta^{}_+ H \eta^{-1}_+\right) P = P^{-1} H^\dagger P = H \; ,
\end{equation}
implying that $[P^{-1} \eta^{}_+, H] = \mathbb{0}_2$. Therefore, we obtain another symmetry operator $C = P^{-1} \eta^{}_+ = \eta^{-1}_+ P = C^{-1}$, which is just the charge-conjugation operator ${\cal C}$ such that $[{\cal C}, {\cal H}] = {\bf 0}$. The existence of a ${\cal C}$ operator for a PT-symmetric non-Hermitian Hamiltonian is necessary to guarantee unitary evolution of the quantum states~\cite{Bender:2002vv}. Furthermore, it is easy to observe that $[{\cal CPT}, {\cal H}] = {\cal C}[{\cal PT}, {\cal H}] + [{\cal C}, {\cal H}]{\cal PT} = {\bf 0}$, indicating that ${\cal CPT}$ is also a symmetry of the Hamiltonian system~\cite{Bender:2007nj}.

\subsection{Inner Products}
\label{sec:IP}

In general, an inner product can be introduced for any two state vectors as follows (see, e.g., Ref.~\cite{Bender:2007nj})
\begin{equation}\label{eq:eta}
\langle \psi|\phi\rangle^{}_\eta \equiv \left(\eta |\psi\rangle\right)^{\rm T} \cdot |\phi\rangle \; ,
\end{equation}
where $\eta$ is an operator in the Hilbert space. A genuine inner product $\langle \psi|\phi\rangle^{}_\eta$ should be a positive-definite, Hermitian, and sesquilinear form. First, the norm $\langle \psi|\psi\rangle^{}_\eta$ of any nonzero vector $|\psi\rangle$ is always a positive real number, and it vanishes only for the zero vector. Second, the inner product $\langle \psi|\phi\rangle^{}_\eta$ is linear in the second vector $|\phi\rangle$, namely, $\langle \psi|c^{}_1 \phi^{}_1 + c^{}_2 \phi^{}_2\rangle^{}_\eta = c^{}_1 \langle \psi|\phi^{}_1\rangle^{}_\eta + c^{}_2 \langle\psi|\phi^{}_2\rangle^{}_\eta$, where $c^{}_{1}, c^{}_{2} \in {\mathbb C}$. Obviously, the Euclidean inner product corresponds to the case of $\eta = {\cal T}$, as indicated in Eq.~(\ref{eq:inner1}). Now, we consider a few alternative definitions of the inner product and examine the norms of the ``mass" and ``flavor" eigenstate vectors. Since the Euclidean inner product is adopted in conventional quantum mechanics, the ${\cal CPT}$ or $\eta^{}_+$ inner products are positive-definite and should be used for non-Hermitian Hamiltonian systems with a PT symmetry~\cite{Mannheim:2017apd}. For this reason, we list the positive-definite inner products below, but collect other possible definitions of inner products in Appendix~\ref{sec:AppB} for completeness.
\begin{itemize}
\item {\bf ${\cal T}$ inner product.} --- As mentioned above, this inner product is equivalent to the Euclidean inner product defined in Eq.~(\ref{eq:inner1}). Using $\eta = {\cal T}$, one can verify that
    \begin{equation}
     \left( \begin{matrix} \langle u^{}_+|u^{}_+\rangle^{}_{\cal T} & \langle u^{}_+|u^{}_-\rangle^{}_{\cal T} \cr \langle u^{}_-|u^{}_+\rangle^{}_{\cal T} & \langle u^{}_-|u^{}_-\rangle^{}_{\cal T}\end{matrix} \right) = \left(\begin{matrix} \sec \alpha & -{\rm i}\tan\alpha \cr {\rm i}\tan\alpha & \sec\alpha \end{matrix}\right) \; ,
    \quad
    \left( \begin{matrix} \langle u^{}_a|u^{}_a\rangle^{}_{\cal T} & \langle u^{}_a|u^{}_b\rangle^{}_{\cal T} \cr \langle u^{}_b|u^{}_a\rangle^{}_{\cal T} & \langle u^{}_b|u^{}_b\rangle^{}_{\cal T}\end{matrix} \right) = \left(\begin{matrix} 1 & 0 \cr 0 & 1\end{matrix}\right) \; .
    \end{equation}
Since the ``flavor'' eigenstate vectors in Eqs.~(\ref{eq:u_a}) and (\ref{eq:u_b}) are real by construction, the ${\cal T}$ inner product is also equivalent to the orthogonal inner product for these vectors.

\item {\bf ${\cal CPT}$ inner product.} --- As we have seen, the charge-conjugate operator ${\cal C}$ commutes with the Hamiltonian ${\cal H}$, i.e., $[{\cal C}, {\cal H}] = {\bf 0}$. The representation matrix for the ${\cal C}$ operator is given by \cite{Kleefeld:2009vd,Bender:2004zz}
    \begin{equation}
    C = P \eta^{}_+ = \eta^{-1}_+ P = \left(\begin{matrix} {\rm i}\tan\alpha & \sec\alpha \cr \sec\alpha & -{\rm i}\tan\alpha \end{matrix}\right)
    \end{equation}
    with $\det C = -1$.
    Thus, we can find the norms with respect to the ${\cal CPT}$ inner product as
    \begin{equation}
    \left( \begin{matrix} \langle u^{}_+|u^{}_+\rangle^{}_{\cal CPT} & \langle u^{}_+|u^{}_-\rangle^{}_{\cal CPT} \cr \langle u^{}_-|u^{}_+\rangle^{}_{\cal CPT} & \langle u^{}_-|u^{}_-\rangle^{}_{\cal CPT}\end{matrix} \right) = \left(\begin{matrix} 1 & 0 \cr 0 & 1\end{matrix}\right)
    \end{equation}
    and
    \begin{equation}
    \left( \begin{matrix} \langle u^{}_a|u^{}_a\rangle^{}_{\cal CPT} & \langle u^{}_a|u^{}_b\rangle^{}_{\cal CPT} \cr \langle u^{}_b|u^{}_a\rangle^{}_{\cal CPT} & \langle u^{}_b|u^{}_b\rangle^{}_{\cal CPT}\end{matrix} \right) =     \left(\begin{matrix} \sec \alpha & -{\rm i}\tan\alpha \cr {\rm i}\tan\alpha & \sec\alpha \end{matrix}\right) \; .
    \end{equation}

\item {\bf $\eta^{}_+$ inner product.} --- In the case of $\eta = \eta^{-1}_+ {\cal T}$, the $\eta^{}_+$ inner product coincides with that in Eq.~(\ref{eq:inner2}). Note that $\eta^{-1}_+ = \eta^{\rm T}_+$. Thus, we can calculate the norms to be
    \begin{equation}
    \left( \begin{matrix} \langle u^{}_+|u^{}_+\rangle^{}_{+} & \langle u^{}_+|u^{}_-\rangle^{}_{+} \cr \langle u^{}_-|u^{}_+\rangle^{}_{+} & \langle u^{}_-|u^{}_-\rangle^{}_{+}\end{matrix} \right) = \left(\begin{matrix} 1 & 0 \cr 0 & 1\end{matrix}\right) \; ,
    \quad
    \left( \begin{matrix} \langle u^{}_a|u^{}_a\rangle^{}_{+} & \langle u^{}_a|u^{}_b\rangle^{}_{+} \cr \langle u^{}_b|u^{}_a\rangle^{}_{+} & \langle u^{}_b|u^{}_b\rangle^{}_{+} \end{matrix} \right) =     \left(\begin{matrix} \sec \alpha & -{\rm i}\tan\alpha \cr {\rm i}\tan\alpha & \sec\alpha \end{matrix}\right)
 \; .
    \end{equation}
\end{itemize}
One can observe that the ${\cal CPT}$ inner product is equivalent to the $\eta^{}_+$ inner product, since $({\cal CP})^{\rm T} = (CP)^{\rm T} = \eta^{}_+ = (\eta^{-1}_+)^{\rm T}$. As remarked in Ref.~\cite{Kleefeld:2009vd}, the ${\cal C}$ operator and thus the metric operator $\eta^{}_+$ are in general not unique for PT-symmetric non-Hermitian Hamiltonians. Therefore, it should be noticed that the equivalence between the ${\cal CPT}$- and $\eta^{}_+$ inner products in the present work, as well as the selection of the matrix $A$ in Eq.~(\ref{eq:A}), should be understood in terms of our convention of normalization for the eigenvectors in Eq.~(\ref{eq:upm}). For the physical Hilbert space, the norm of any state vector is conventionally assumed to be positive and a zero norm is obtained only for the null state vector. On the other hand, the ${\cal P}$ operator has eigenvalues $\pm 1$ and is not positive-definite. Therefore, only the ${\cal CPT}$ and $\eta^{}_+$ inner products are appropriate choices, which are actually equivalent. As we have already observed, the ${\cal T}$, ${\cal CPT}$, and $\eta^{}_+$ inner products imply that $\eta$ in Eq.~(\ref{eq:eta}) is a linear positive-definite operator multiplied by the time-reversal operator ${\cal T}$.

\subsection{Transition Amplitudes and Probabilities}

As in conventional quantum mechanics, one assumes that the time evolution of the ``mass" eigenstate vectors is dictated by the Schr\"{o}dinger equation, which can be easily solved as
\begin{equation}
|u^{}_\pm(t) \rangle = e^{- {\rm i} E^{}_\pm t} |u^{}_\pm\rangle \; ,
\end{equation}
where the state vectors without arguments are just those at time $t = 0$, namely, $|u^{}_\pm\rangle \equiv |u^{}_\pm(0)\rangle$. Hence, the time evolution of the ``flavor" eigenstate vectors is governed by
\begin{eqnarray}
|u^{}_{a}(t)\rangle &=& \left(A^{-1}\right)^{}_{a+} e^{-{\rm i}E^{}_+ t}|u^{}_+\rangle + \left(A^{-1}\right)^{}_{a-} e^{-{\rm i}E^{}_- t} |u^{}_-\rangle \; , \\
|u^{}_{b}(t)\rangle &=& \left(A^{-1}\right)^{}_{b+} e^{-{\rm i}E^{}_+ t} |u^{}_+\rangle +\left(A^{-1}\right)^{}_{b-} e^{-{\rm i}E^{}_- t} |u^{}_-\rangle \; .
\end{eqnarray}
An immediate question is how to define the transition amplitude and the corresponding transition probability. For a general inner product $\langle \psi|\phi\rangle^{}_\eta$, where $\eta$ is an operator in the Hilbert space, we define the transition amplitude
\begin{equation}\label{eq:amplitudeold}
{\cal A}^\eta_{\alpha \beta} \equiv \langle u^{}_\beta | u^{}_\alpha(t)\rangle^{}_\eta
\end{equation}
and the corresponding transition probability
\begin{equation}\label{eq:probold}
{\cal P}^\eta_{\alpha \beta} \equiv \frac{\left|\langle u^{}_\beta | u^{}_\alpha(t)\rangle^{}_\eta\right|^2}{\langle u^{}_\beta|u^{}_\beta\rangle^{}_\eta \langle u^{}_\alpha(t)|u^{}_\alpha(t)\rangle^{}_\eta} \; ,
\end{equation}
where the $\eta$ inner product should be properly chosen to ensure that the norm is time-independent as in conventional quantum mechanics. As mentioned before, the ${\cal CPT}$ or $\eta^{}_+$ inner products could be adopted such that the norm of a nonzero vector is always a positive and real number. Only, in this way, the denominator on the right-hand side of Eq.~(\ref{eq:probold}) can be positive and the transition probabilities physically meaningful.

If we adopt the ${\cal CPT}$ inner product, as defined in Subsection~\ref{sec:IP}, it is straightforward to calculate the absolute value squared of the transition amplitudes as
\begin{eqnarray}
\left|{\cal A}^{\cal CPT}_{aa}\right|^2 &=& \sec^2\alpha \cos^2 \frac{\beta t}{2} \; , \\
\left|{\cal A}^{\cal CPT}_{ab}\right|^2 &=& \sec^2\alpha \sin^2 \left(\alpha - \frac{\beta t}{2}\right) \; , \\
\left|{\cal A}^{\cal CPT}_{ba}\right|^2 &=& \sec^2\alpha \sin^2 \left(\alpha + \frac{\beta t}{2}\right) \; , \\
\left|{\cal A}^{\cal CPT}_{bb}\right|^2 &=& \sec^2\alpha \cos^2 \frac{\beta t}{2} \; ,
\end{eqnarray}
where $\beta \equiv E^{}_+ - E^{}_- = 2\sqrt{\sigma^2 - \rho^2 \sin^2\varphi}$ is the energy eigenvalue difference between two ``mass" eigenstates. Given the ${\cal CPT}$ norms of the ``flavor'' eigenstates $\langle u^{}_a|u^{}_a\rangle^{}_{\cal CPT} = \langle u^{}_b|u^{}_b\rangle^{}_{\cal CPT} = \sec \alpha$ and the corresponding time evolutions $\langle u^{}_a(t)|u^{}_a(t)\rangle^{}_{\cal CPT} = \langle u^{}_b(t)|u^{}_b(t)\rangle^{}_{\cal CPT} = \sec \alpha$ (which are time-independent), it is easy to verify that
\begin{eqnarray}
{\cal P}^{\cal CPT}_{aa} + {\cal P}^{\cal CPT}_{ab} &=& \frac{1}{2} \left[2 + \cos \beta t - \cos (2\alpha + \beta t)\right] \; , \\
{\cal P}^{\cal CPT}_{ba} + {\cal P}^{\cal CPT}_{bb} &=& \frac{1}{2} \left[2 + \cos \beta t - \cos (2\alpha - \beta t)\right] \; ,
\end{eqnarray}
implying that the probability is unfortunately not conserved. Even for $t = 0$, one can observe that ${\cal P}^{\cal CPT}_{aa} + {\cal P}^{\cal CPT}_{ab} = {\cal P}^{\cal CPT}_{ba} + {\cal P}^{\cal CPT}_{bb} = 1 + \sin^2\alpha \neq 1$. This observation can be ascribed to the non-unitary mixing matrix $A^{-1}$ in Eq.~(\ref{eq:A}) that connects the ``mass" eigenstates $\{|u^{}_+\rangle, |u^{}_-\rangle\}$ and the ``flavor" eigenstates $\{|u^{}_a\rangle, |u^{}_b\rangle\}$. (Cf.~the discussion about neutrino oscillations and the zero-distance effect later in this work.)

It is well known that the pseudo-Hermitian Hamiltonian $H$ in Eq.~(\ref{eq:Ham}) can be converted into a Hermitian one via the similarity transformation $G H G^{-1} = H^\prime$, where the Hermitian matrix $G$ is given by~\cite{Mostafazadeh:2002id, Mostafazadeh:2008pw, Mannheim:2009zj}
\begin{eqnarray}\label{eq:G}
G = \frac{1}{\sqrt{\cos \alpha}} \left(\begin{matrix} \displaystyle \cos \frac{\alpha}{2} & \displaystyle -{\rm i} \sin \frac{\alpha}{2} \cr \displaystyle {\rm i}\sin \frac{\alpha}{2} & \displaystyle \cos \frac{\alpha}{2}\end{matrix}\right) \; ,
\end{eqnarray}
which is related to the metric matrix $\eta^{}_+$ via $G^2 = \eta^{}_+$. Furthermore, it is straightforward to verify that $G^{-1} = G^* = G^{\rm T}$. The Hermitian counterpart of $H$ turns out to be real and symmetric, i.e.,
\begin{eqnarray}\label{eq:HermitianH}
H^\prime = \left(\begin{matrix} \rho \cos\varphi & \sqrt{\sigma^2 - \rho^2 \sin^2\varphi} \cr \sqrt{\sigma^2 - \rho^2 \sin^2 \varphi} & \rho \cos\varphi\end{matrix}\right) \; ,
\end{eqnarray}
which can be diagonalized via $V H^\prime V^{-1} = \widehat{H}$ with the orthogonal matrix
\begin{eqnarray}\label{eq:V}
V = \frac{1}{\sqrt{2}}\left(\begin{matrix} 1 & 1 \cr 1 & -1 \end{matrix}\right) \; .
\end{eqnarray}
The diagonal matrix $\widehat{H} \equiv {\rm diag}(E^{}_+, E^{}_-)$ with $E^{}_\pm = \rho \cos\varphi \pm \sqrt{\sigma^2 - \rho^2 \sin^2 \varphi}$ remains to be the same, as it should. The corresponding eigenvectors are then given as
\begin{eqnarray}\label{eq:neweigenstates}
|u^\prime_\pm\rangle &=& \frac{1}{\sqrt{2}} \left(\begin{matrix} 1 \cr \pm 1 \end{matrix}\right) \; ,
\end{eqnarray}
where the proper phase convention has been chosen such that $A = V G$ can be identified.

Using the Hermitian Hamiltonian $H^\prime$, one can now apply conventional quantum mechanics to describe the time evolution of any quantum states. Therefore, it is interesting to make a connection between the observables in both non-Hermitian and Hermitian quantum mechanics. Due to the transformation of the Hamiltonian $H^\prime = G H G^{-1}$, we have to accordingly redefine the ``flavor" eigenstate vectors by
\begin{eqnarray}\label{eq:newflavor}
|u^\prime_\alpha\rangle \equiv G|u^{}_\alpha\rangle
\end{eqnarray}
and recast the transition amplitudes and probabilities in Eqs.~(\ref{eq:amplitudeold}) and (\ref{eq:probold}) with the $\eta^{}_+$ inner product into
\begin{eqnarray}\label{eq:ampnew}
{\cal A}^{+}_{\alpha\beta} \equiv \langle u^{}_\beta|u^{}_\alpha(t)\rangle^{}_{+} = \langle u^\prime_\beta| G^{-1} \eta^{}_+ G^{-1}|u^\prime_\alpha(t)\rangle = \langle u^\prime_\beta|u^\prime_\alpha(t)\rangle
\end{eqnarray}
and
\begin{eqnarray}
{\cal P}^{+}_{\alpha\beta} \equiv \frac{\left|\langle u^{}_\beta | u^{}_\alpha(t)\rangle^{}_{+}\right|^2}{\langle u^{}_\beta|u^{}_\beta\rangle^{}_{+} \langle u^{}_\alpha(t)|u^{}_\alpha(t)\rangle^{}_{+}} = \frac{\left| \langle u^\prime_\beta|u^\prime_\alpha(t)\rangle \right|^2}{\langle u^\prime_\beta|u^\prime_\beta\rangle \langle u^\prime_\alpha(t)|u^\prime_\alpha(t)\rangle} \; ,
\end{eqnarray}
where the identity $G^{-1} \eta^{}_+ G^{-1} = \mathbb{1}_2$ has been used. Hence, it becomes clear that the transition amplitudes and probabilities should be identical if the time evolution of $|u^\prime_\alpha(t)\rangle$ is governed by the Hermitian Hamiltonian $H^\prime$ and the Euclidean inner product is used as in conventional quantum mechanics.

Since the ${\cal CPT}$ inner product is equivalent to the $\eta^{}_+$ inner product, our discussion around Eqs.~(\ref{eq:prob2}) and (\ref{eq:probtime}) in the latter case are also applicable to the former one. In the two-level quantum system with the Hamiltonian $H$ in Eq.~(\ref{eq:Ham}), one can show that the total probability for the transitions between two ``flavor" eigenstates is conserved only if the Hamiltonian is Hermitian. A possible solution to the problem of probability non-conservation could be to introduce an auxiliary flavor state~\footnote{Unitary evolution of the state vectors is guaranteed in a PT-symmetric non-Hermitian Hamiltonian system, whereas conservation of the total probability for the transitions between two ``flavor" eigenstates is not. The introduction of the ${\cal CPT}$ eigenstates may serve as a solution to this problem.}
\begin{equation}
|\widetilde{u}^{}_\alpha\rangle \equiv \frac{1}{2} \left(|u^{}_\alpha\rangle + {\cal CPT}|u^{}_\alpha\rangle \right) \; ,
\end{equation}
which is actually a direct eigenstate of the ${\cal CPT}$ operator, i.e., ${\cal CPT}|\widetilde{u}^{}_\alpha\rangle = |\widetilde{u}^{}_\alpha\rangle$ (for $\alpha = a, b$). Requiring the final state in the transition process to be the newly-defined flavor state, we can again compute the transition amplitudes ${\cal A}^{\cal CPT}_{\alpha \beta} \equiv \langle \widetilde{u}^{}_\beta|u^{}_\alpha(t)\rangle^{}_{\cal CPT}$ and the corresponding transition probabilities
\begin{equation}
{\cal P}^{\cal CPT}_{\alpha \beta} \equiv \frac{\left|\langle \widetilde{u}^{}_\beta|u^{}_\alpha(t)\rangle^{}_{\cal CPT}\right|^2}{\langle \widetilde{u}^{}_\beta|\widetilde{u}^{}_\beta\rangle^{}_{\cal CPT} \langle u^{}_\alpha(t)|u^{}_\alpha(t)\rangle^{}_{\cal CPT}} \; .
\end{equation}
Since the norms $\langle \widetilde{u}^{}_\beta|\widetilde{u}^{}_\beta\rangle^{}_{\cal CPT}$ and $\langle u^{}_\alpha(t)|u^{}_\alpha(t)\rangle^{}_{\cal CPT}$ are time-independent, it is only necessary to care about the time dependence of the summation of the absolute value squared transition amplitudes over the final states. More explicitly, we obtain
\begin{equation}\label{eq:sumA}
\sum_{\beta=a,b} \left|{\cal A}^{\cal CPT}_{\alpha \beta}\right|^2 = \sum_\beta \langle u^{}_\alpha(t)|\widetilde{u}^{}_\beta\rangle^{}_{\cal CPT} \langle \widetilde{u}^{}_\beta|u^{}_\alpha(t)\rangle^{}_{\cal CPT} = \frac{1}{4} \langle u^{}_\alpha(t)|(1 + 2\eta^{}_+ + \eta^2_+)|u^{}_\alpha(t)\rangle^{}_{\cal T} \; ,
\end{equation}
where the identity $({\cal CP})^{\rm T} = \eta^{}_+$ has been implemented. Given $\eta^{-1}_+$ in Eq.~(\ref{eq:complete}) and $\eta^{-1}_+ = \eta^{\rm T}_+$, one can prove that $1 + \eta^2_+ = 2 \eta^{}_+ \sec\alpha$. Therefore, Eq.~(\ref{eq:sumA}) can be rewritten as
\begin{equation}\label{eq:sumA2}
\sum_\beta \left|{\cal A}^{\cal CPT}_{\alpha \beta}\right|^2 =  \langle u^{}_\alpha(t)|u^{}_\alpha(t)\rangle^{}_+ \sec\alpha \cos^2 \frac{\alpha}{2} = \sec^2\alpha \cos^2\frac{\alpha}{2}\; ,
\end{equation}
which is evidently time-independent. Note that $\langle u^{}_a|u^{}_a\rangle^{}_+ = \langle u^{}_b|u^{}_b\rangle^{}_+ = \sec\alpha$ has been implemented in the second equality in Eq.~(\ref{eq:sumA2}). Therefore, the individual transition probabilities are given by
\begin{eqnarray}
{\cal P}^{\cal CPT}_{aa}  &=& \cos^2 \frac{\alpha - \beta t}{2} \; , \label{eq:Paa}\\
{\cal P}^{\cal CPT}_{ab}  &=& \sin^2 \frac{\alpha - \beta t}{2} \; , \label{eq:Pab}\\
{\cal P}^{\cal CPT}_{ba}  &=& \sin^2 \frac{\alpha + \beta t}{2} \; , \\
{\cal P}^{\cal CPT}_{bb}  &=& \cos^2 \frac{\alpha + \beta t}{2} \; ,
\end{eqnarray}
where one can observe that
\begin{equation}
\frac{1}{2} \left( {\cal P}^{\cal CPT}_{aa} + {\cal P}^{\cal CPT}_{ab} + {\cal P}^{\cal CPT}_{ba} + {\cal P}^{\cal CPT}_{bb} \right) = 1 \; ,
\end{equation}
which serves as a ``weak'' condition for conservation of probability. Furthermore, one can observe that ${\cal P}^{\cal CPT}_{aa} + {\cal P}^{\cal CPT}_{ab} = 1$ and ${\cal P}^{\cal CPT}_{ba} + {\cal P}^{\cal CPT}_{bb} = 1$, which are ``strong'' conditions for conservation of probability. However, ${\cal P}^{\cal CPT}_{aa} + {\cal P}^{\cal CPT}_{ba} = 1 + \sin \alpha \sin \beta t \neq 1$ and ${\cal P}^{\cal CPT}_{ab} + {\cal P}^{\cal CPT}_{bb} = 1 - \sin \alpha \sin \beta t \neq 1$ are not conserved and even time-dependent. Hence, the time-reversal asymmetry $\Delta_{\cal T} \equiv {\cal P}^{\cal CPT}_{ab} - {\cal P}^{\cal CPT}_{ba} = {\cal P}^{\cal CPT}_{bb} - {\cal P}^{\cal CPT}_{aa}$ is given by $
\Delta_{\cal T} = -\sin \alpha \sin \beta t$.
Finally, applying the ${\cal C}$ operator on the ``flavor'' eigenstates $|u_\alpha\rangle$, one can define ``antiflavor'' eigenstates as $|\overline{u}_\alpha\rangle \equiv {\cal C} |u_\alpha\rangle$, which implies that $\langle \overline{\widetilde{u}}_\beta | \overline{u}_\alpha(t) \rangle_{\cal CPT} =  \langle \widetilde{u}_\beta | u_\alpha(t) \rangle_{\cal CPT}$, so that in turn ${\cal A}^{\cal CPT}_{\bar{\alpha} \bar{\beta}} = {\cal A}^{\cal CPT}_{\alpha \beta}$ and ${\cal P}^{\cal CPT}_{\bar{\alpha} \bar{\beta}} = {\cal P}^{\cal CPT}_{\alpha \beta}$. Thus, the ${\cal CPT}$ asymmetry $\Delta_{\cal CPT} \equiv {\cal P}^{\cal CPT}_{ab} - {\cal P}^{\cal CPT}_{\bar{b}\bar{a}} = -({\cal P}^{\cal CPT}_{ba} - {\cal P}^{\cal CPT}_{\bar{a}\bar{b}}) = - \sin \alpha \sin \beta t$, but it holds that ${\cal P}^{\cal CPT}_{\bar{a}\bar{a}} = {\cal P}^{\cal CPT}_{aa}$ and ${\cal P}^{\cal CPT}_{\bar{b}\bar{b}} = {\cal P}^{\cal CPT}_{bb}$. In fact, note that $\Delta_{\cal CPT} = \Delta_{\cal T}$.

In summary, it is reasonable to introduce ${\cal CPT}$ eigenstates $|\widetilde{u}^{}_\alpha\rangle = \left(|u^{}_\alpha\rangle + {\cal CPT}|u^{}_\alpha\rangle\right)/2$ (for $\alpha = a,b$) and interpret ${\cal A}^{\cal CPT}_{\alpha \beta} \equiv \langle \widetilde{u}^{}_\beta|u^{}_\alpha(t)\rangle^{}_{\cal CPT}$ (for $\alpha,\beta = a,b$) as the transition amplitudes, which lead to the transition probabilities that satisfy conservation of probability. One may wonder if it is possible to achieve conservation of probability given the transition amplitudes $\langle \widetilde{u}^{}_\beta|\widetilde{u}^{}_\alpha(t)\rangle^{}_{\cal CPT}$, where the system is prepared initially in the ${\cal CPT}$ eigenstates $|\widetilde{u}^{}_\alpha\rangle$ and then evolved to time $t$ as $|\widetilde{u}^{}_\alpha(t)\rangle$. Unfortunately, we have verified that this does not work. Since $|u^{}_+\rangle$ and $|u^{}_-\rangle$ are also eigenstates of the ${\cal CPT}$ operator, one can verify that $|\widetilde{u}^{}_a\rangle$ and $|\widetilde{u}^{}_b\rangle$ are related to $|u^{}_+\rangle$ and $|u^{}_-\rangle$ by an orthogonal matrix with a maximal rotation angle.

\section{Two-Flavor Neutrino Oscillations}\label{sec:3}

Now, we apply the formalism in Sec.~\ref{sec:2} to the case of two-flavor neutrino oscillations. As is well known, the Hamiltonian for neutrino oscillations in vacuum reads
\begin{equation}\label{eq:H2nu}
{\cal H}^{}_{\rm vac} = \frac{1}{2E} \left( \begin{matrix} \cos \theta^{}_{23} & \sin \theta^{}_{23} \cr -\sin \theta^{}_{23} & \cos \theta^{}_{23} \end{matrix} \right) \left( \begin{matrix} m^2_2 & 0 \cr 0 & m^2_3 \end{matrix} \right) \left( \begin{matrix} \cos \theta^{}_{23} & -\sin \theta^{}_{23} \cr \sin \theta^{}_{23} & \cos \theta^{}_{23} \end{matrix} \right) \; ,
\end{equation}
where $\theta^{}_{23}$ is the leptonic mixing angle and $m^2_i$ (for $i = 2, 3$) are the neutrino masses in vacuum.The choice of the parameters $\theta_{23}$ and $\Delta m_{32}^2 \equiv m_3^2 - m_2^2$ is motivated by atmospheric neutrino oscillations (which are governed by these parameters) discovered by the Super-Kamiokande experiment in 1998 \cite{Fukuda:1998mi}, which shows that neutrinos have mass. The present $3\sigma$ ranges of these parameters are $41^\circ \lesssim \theta_{23} \lesssim 52^\circ$ and $2.4 \times 10^{-3} \, {\rm eV^2} \lesssim |\Delta m_{32}^ 2| \lesssim 2.6 \times 10^{-3} \, {\rm eV^2}$ \cite{Esteban:2018azc}. Still, the value of $\theta_{23}$ is compatible with that of maximal mixing, i.e., $\theta_{23} = 45^\circ$. The Hamiltonian ${\cal H}^{}_{\rm vac}$ in Eq.~(\ref{eq:H2nu}) is obviously Hermitian, but not PT-symmetric, i.e., $[{\cal PT}, {\cal H}^{}_{\rm vac}] \neq {\bf 0}$. However, for $\theta^{}_{23} = \pi/4$, one can prove that ${\cal H}^{}_{\rm vac}$ is also PT-symmetric. Therefore, the most general PT-symmetric non-Hermitian Hamiltonian relevant for two-flavor neutrino oscillations is
\begin{equation}\label{eq:H2nuPT}
{\cal H} = \frac{1}{4E} \left[ \left( \begin{matrix} m^2_2 + m^2_3 & \Delta m^2_{32} \cr \Delta m^2_{32} & m^2_2 + m^2_3 \end{matrix} \right) + \left(\begin{matrix} \rho e^{{\rm i}\varphi} & \sigma e^{{\rm i}\phi} \cr  \sigma e^{-{\rm i}\phi} & \rho e^{-{\rm i}\varphi} \end{matrix}\right) \right] \; ,
\end{equation}
where $\Delta m^2_{32} \equiv m^2_3 - m^2_2$ and the second term is PT-symmetric. If we further require the Hamiltonian to be symmetric, the phase $\phi$ should be set to zero. In general, the PT-symmetric non-Hermitian correction matrix, i.e., the second term in Eq.~(\ref{eq:H2nuPT}), can be non-symmetric, but this symmetric form is assumed in accordance with our previous discussion on the Hamiltonian $H$ in Eq.~(\ref{eq:Ham}). Now, remember that ${\cal T} $ has the property ${\cal T}^2 = 1$. However, neutrinos are fermions with spin 1/2, and therefore, as argued in Ref.~\cite{JonesSmith:2009wy}, ${\cal T}$ should instead fulfill ${\cal T}^2 = -1$. Nevertheless, for neutrino oscillations, neutrinos can be considered particles governed by a Schr{\"o}dinger-like equation instead of the Dirac equation. Thus, in the ultra-relativistic limit, which is valid for neutrino oscillations in all realistic experiments, the dynamics of neutrinos described by the Dirac equation is reduced to the time evolution of neutrino flavor eigenstates governed by the Schr{\"o}dinger equation that does not reveal the full spin structure of neutrinos. Hence, for neutrino oscillations in the ultra-relativistic limit, neutrinos behave as spinless particles, and therefore, it can be assumed that ${\cal T}$ fulfills ${\cal T}^2 = 1$, which is adopted in the construction of the Hamiltonian $H$ in Eq.~(\ref{eq:Ham}). The Hamiltonian ${\cal H}$ in Eq.~(\ref{eq:H2nuPT}) can then be rewritten as
\begin{eqnarray}\label{eq:H2nuPT0}
{\cal H} &=& \frac{1}{4E} \left[ \left( \begin{matrix} \overline{m}^2 & \Delta m^2_{32} \cr \Delta m^2_{32} & \overline{m}^2 \end{matrix} \right) + \left(\begin{matrix} \rho e^{{\rm i}\varphi} & \sigma \cr  \sigma & \rho e^{-{\rm i}\varphi} \end{matrix}\right) \right] \nonumber \\
&=& \frac{1}{4E} \left( \begin{matrix} \overline{m}^2 + \rho \cos\varphi + {\rm i}\rho\sin\varphi & \Delta m^2_{32} + \sigma \cr \Delta m^2_{32} + \sigma & \overline{m}^2 + \rho \cos\varphi - {\rm i}\rho\sin\varphi \end{matrix} \right) \; ,
\end{eqnarray}
where $\overline{m}^2 \equiv m^2_2 + m^2_3$. In the following, we concentrate on the two-level quantum system governed by the Hamiltonian ${\cal H}$ in Eq.~(\ref{eq:H2nuPT0}) and examine its implications for two-flavor neutrino oscillations in vacuum. In the limit of $\varphi = 0$, the Hermiticity of the Hamiltonian ${\cal H}$ is recovered. Now, some useful observations can be made.
\begin{itemize}
\item The eigenvalues of the Hamiltonian ${\cal H}$ can be easily figured out to be
\begin{equation}
\omega^{}_\pm = \frac{1}{4E} \left( \overline{m}^2 + \rho \cos\varphi \pm \sqrt{(\Delta m^2_{32} + \sigma)^2 - \rho^2 \sin^2\varphi } \right) \; ,
\end{equation}
where $|\rho\sin\varphi| \leq |\Delta m^2_{32} + \sigma|$ should be satisfied to respect the exact PT symmetry. The corresponding eigenvectors are given by [cf., Eq.~(\ref{eq:upm})]
\begin{equation}
|\nu^{}_\pm\rangle =  \frac{1}{\sqrt{2\cos \alpha^\prime}} \left(\begin{matrix} e^{\pm {\rm i}\alpha^\prime/2} \cr \pm e^{\mp {\rm i}\alpha^\prime/2}\end{matrix}\right) \; ,
\end{equation}
where $\sin\alpha^\prime \equiv \rho \sin\varphi/(\Delta m^2_{32} + \sigma)$. The neutrino mass eigenstate basis $\{|\nu^{}_+\rangle, |\nu^{}_-\rangle\}$ is related to the neutrino flavor eigenstate basis $\{|\nu^{}_\mu\rangle, |\nu^{}_\tau\rangle\}$ by the ``mixing" matrix $A^{-1}$ as in Eq.~(\ref{eq:A}), but with $\alpha$ replaced by $\alpha^\prime$, which is not unitary.

\item As discussed in Sec.~\ref{sec:2}, we can calculate the transition probabilities for $|\nu^{}_\mu\rangle \to |\widetilde{\nu}^{}_\mu\rangle$ and $|\nu^{}_\mu \rangle \to |\widetilde{\nu}^{}_\tau\rangle$ by introducing the ${\cal CPT}$ eigenstates
    \begin{eqnarray}
    |\widetilde{\nu}^{}_\mu\rangle &=& \frac{1}{2} \left(|\nu^{}_\mu\rangle + {\cal CPT}|\nu^{}_\mu\rangle\right) = \frac{\cos (\alpha^\prime/2)}{\sqrt{2\cos\alpha^\prime}} \left(|\nu^{}_+\rangle + |\nu^{}_-\rangle\right) \; , \\
    |\widetilde{\nu}^{}_\tau\rangle &=& \frac{1}{2} \left(|\nu^{}_\tau\rangle + {\cal CPT}|\nu^{}_\tau\rangle\right) = \frac{\cos (\alpha^\prime/2)}{\sqrt{2\cos\alpha^\prime}} \left(|\nu^{}_+\rangle - |\nu^{}_-\rangle\right) \; .
    \end{eqnarray}
    The transition probabilities can be readily read off from Eqs.~(\ref{eq:Paa}) and (\ref{eq:Pab}), viz.,
    \begin{eqnarray}
    {\cal P}^{}_{\mu\mu} &=& 1 - \sin^2 \frac{\alpha^\prime - \beta^\prime t}{2} = 1 - \sin^2 \frac{\beta^\prime t}{2} + \frac{1}{2} \sin (\beta^\prime t) \, \alpha^\prime + {\cal O}({\alpha^\prime}^2) \; , \label{eq:Pmumu}\\
    {\cal P}^{}_{\mu\tau} &=& \sin^2 \frac{\alpha^\prime - \beta^\prime t}{2} = \sin^2 \frac{\beta^\prime t}{2} - \frac{1}{2} \sin (\beta^\prime t) \, \alpha^\prime + {\cal O}({\alpha^\prime}^2) \; , \label{eq:Pmutau}
    \end{eqnarray}
    where $\beta^\prime \equiv \omega_+ - \omega_- = \sqrt{(\Delta m^2_{32} + \sigma)^2 - \rho^2 \sin^2\varphi}/(2E)$. Compared to the ordinary two-flavor neutrino oscillation probabilities ${\cal P}^{}_{\mu\mu} = 1 - \sin^2[\Delta m^2_{32} \, t/(4E)]$ and ${\cal P}^{}_{\mu\tau} = \sin^2[\Delta m^2_{32} \, t/(4E)]$, the ones in Eqs.~(\ref{eq:Pmumu}) and (\ref{eq:Pmutau}) reveal two main features. First, the non-Hermitian parameter $\alpha^\prime$ appears in the oscillation phase shift, resulting in a so-called zero-distance effect \cite{Langacker:1988up}. Second, the oscillation phase in the non-Hermitian case will be reduced to that in the standard case if the non-Hermitian parameters $\rho$, $\sigma$, and $\varphi$ are switched off. Since $\theta^{}_{23} \approx \pi/4$ is still consistent with all neutrino oscillation experiments, it may hint at a non-Hermitian description. At least, future high-precision measurements of neutrino oscillation probabilities will be able to constrain the non-Hermitian and PT-symmetric corrections.

\begin{figure}[!t]
\begin{center}
\includegraphics[width=0.8\textwidth]{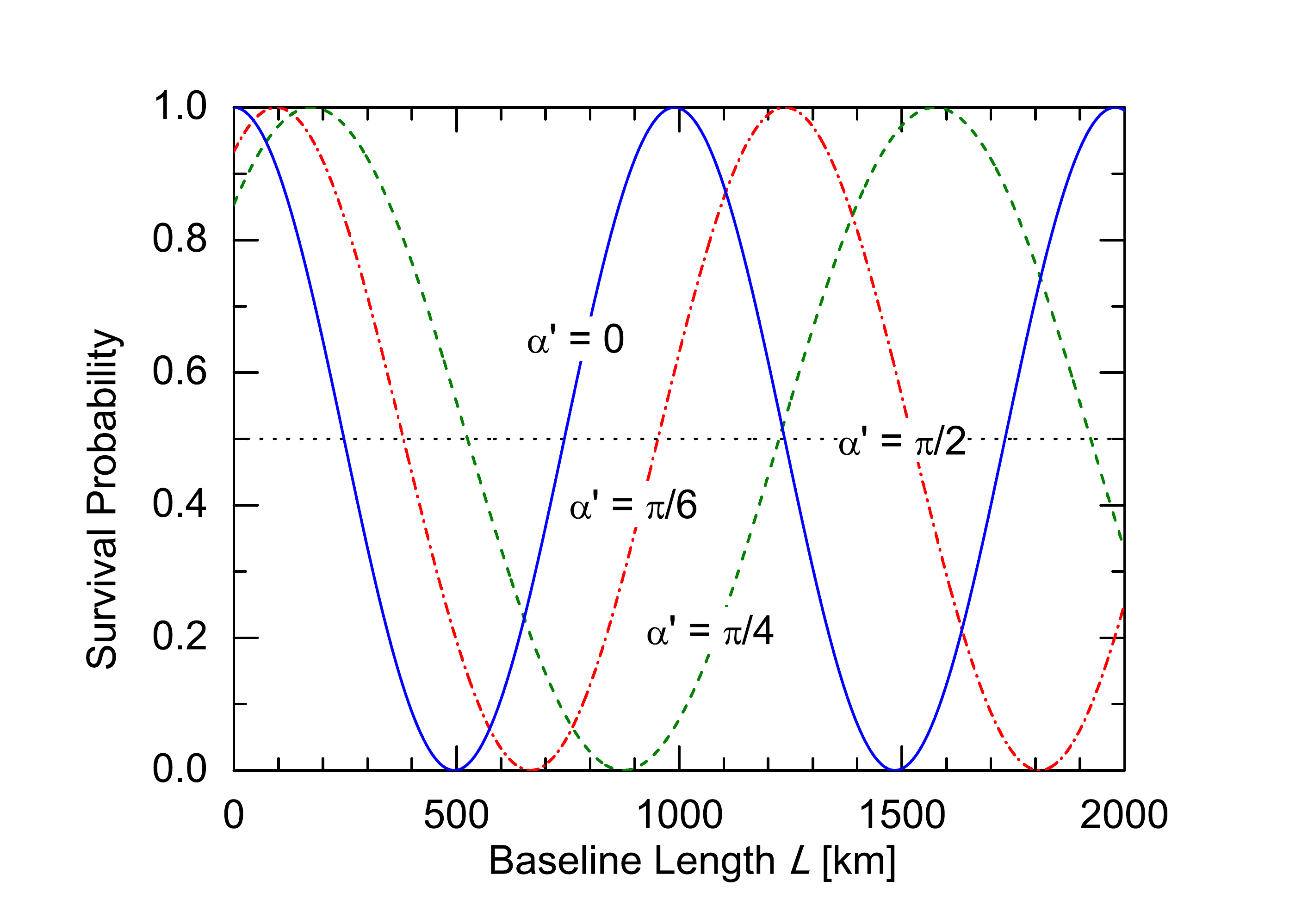}
\end{center}
\vspace{-0.6cm}
\caption{Survival transition probability ${\cal P}_{\mu\mu}$ as a function of the baseline length $L = ct$ with $c \simeq 3\times 10^{8} \, {\rm m/s}$ being the speed of light. The blue solid curve shows the ordinary Hermitian case with $\alpha^\prime = 0$. The red dash-dotted curve is for $\alpha^\prime = \pi/6$, whereas the green dashed curve is for $\alpha^\prime = \pi/4$. The black dotted horizontal line shows the case for $\alpha^\prime = \pi/2$. The assumed values of the other parameters are $\Delta m^2_{32} = 2.5\times 10^{-3} \, {\rm eV}^2$, $\sigma = 0$, and $E = 1 \, {\rm GeV}$.}
\label{fig:survprob}
\end{figure}

In Fig.~1, we display the survival  probability ${\cal P}_{\mu\mu}$ given in Eq.~(\ref{eq:Pmumu}) for two-flavor neutrino oscillations in vacuum with $\nu_\mu$ and $\nu_\tau$ using four different values of the oscillation phase shift parameter $\alpha^\prime \in \{0,\pi/6,\pi/4,\pi/2\}$. To better interpret these results, we rewrite Eq.~(\ref{eq:Pmumu}) in a more useful form
\begin{eqnarray}
{\cal P}^{}_{\mu\mu} &=& 1 - \sin^2 \left( \frac{\alpha^\prime}{2} - \frac{\beta^\prime}{2} \; t \right) = 1 - \sin^2 \left(\frac{\alpha^\prime}{2} - \frac{\sqrt{(\Delta m^2_{32} + \sigma)^2 - \rho^2 \sin^2\varphi}}{4E} \; t \right) \nonumber \\
&\simeq& 1 - \sin^2 \left[\frac{\alpha^\prime}{2} - 1.27 \cos\alpha^\prime \cdot \left(\frac{\Delta m^2_{32} + \sigma}{{\rm eV}^2}\right) \left(\frac{\rm GeV}{E}\right) \left(\frac{L}{\rm km}\right)\right] \; ,
\end{eqnarray}
where $\beta^\prime = (\Delta m^2_{32} + \sigma)\cos \alpha^\prime/(2E)$ and $\cos\alpha^\prime = \sqrt{1 - \rho^2\sin^2\varphi/(\Delta m^2_{32} + \sigma)^2}$ have been used and the time $t$ has been identified with the baseline length $L = ct$ of neutrino propagation with $c \simeq 3 \times 10^{8} \, {\rm m/s}$ being the speed of light. For illustration, $\Delta m^2_{32} = 2.5\times 10^{-3}~{\rm eV}^2$, $\sigma = 0$, and $E = 1~{\rm GeV}$ have been assumed in producing the results of Fig.~1, where we observe how the different values of $\alpha'$ shift the oscillation pattern (with maintained maximal oscillation amplitude) from left to right.\footnote{In order to ensure PT symmetry, the two non-Hermtian parameters $\rho$ and $\varphi$ must fulfill the condition $|\rho \sin \varphi | \leq 2.5\times 10^{-3}~{\rm eV}^2$, otherwise the PT symmetry will be broken.} For $\alpha^\prime = 0$, we have the ordinary survival probability ${\cal P}_{\mu\mu}(L) = 1 - \sin^2[\Delta m^2_{32}L/(4E)]$ and no zero-distance effect. Then, for $\alpha^\prime = \pi/6$ and $\alpha^\prime = \pi/4$, the oscillation phase is shifted from $\Delta m^2_{32}L/(4E)$ to $(\alpha^\prime - \beta^\prime L)/2 = \alpha^\prime/2 - \cos\alpha^\prime \Delta m^2_{32} L/(4E)$, leading to a larger zero-distance effect for an increasing value of $\alpha^\prime$. Meanwhile, it should be noticed that, in addition to a constant phase shift of $\alpha^\prime/2$, the oscillation frequency is suppressed by a factor of $\cos \alpha^\prime$. Although we have set $\sigma = 0$ in our calculations, one can easily observe that a nonzero value of $\sigma$ would lead to a change in the oscillation frequency, mimicking a different value of $\Delta m^2_{32}$. Finally, for $\alpha^\prime = \pi/2$, the survival probability collapses to ${\cal P}_{\mu\mu}(L) = 1 - \sin^2(\pi/4) = 1/2$ and the zero-distance effect is half of the oscillation amplitude. In this case, the oscillation pattern simply disappears, since $\cos\alpha^\prime = 0$.
\end{itemize}

The realistic picture of neutrino oscillations involves three neutrino flavors. Thus, one may generalize the matrix representation of the parity operator ${\cal P}$ in Eq.~(\ref{eq:Pmatrx}) to a $3\times 3$ matrix, which can be chosen to realize the exchange of $\mu$ and $\tau$ neutrino flavors. When matter effects on neutrino oscillations are taken into account, the exact PT symmetry can be preserved for an arbitrary mixing angle, but the Mikheyev--Smirnov--Wolfenstein resonance condition \cite{Wolfenstein:1977ue,Mikheev:1986gs,Mikheev:1986wj} has to be satisfied~\cite{Ohlsson:2015xsa}. These generalizations deserve further investigations.

\section{Summary and Conclusions}\label{sec:4}

Motivated by the recent tremendous progress in conceptual understanding and practical applications of PT-symmetric non-Hermitian Hamiltonians, we have examined how to consistently define the transition amplitudes ${\cal A}^{}_{\alpha \beta}$ and probabilities ${\cal P}^{}_{\alpha \beta} = |{\cal A}^{}_{\alpha \beta}|^2$ for ``flavor" transitions $|u^{}_\alpha\rangle \to |u^{}_\beta\rangle$ in a finite-dimensional non-Hermitian quantum system, where $|u^{}_\alpha\rangle$ and $|u^{}_\beta\rangle$ for $\alpha, \beta = 1, 2, \cdots, N$ stand for the ``flavor" eigenstate vectors and $N$ is the dimension of the Hilbert space. The general criterion for consistency is assumed to be conservation of transition probability. In other words, the sum of the transition probabilities over the final states $\sum^N_{\beta=1} {\cal P}^{}_{\alpha \beta}$ should be time-independent. In the simplest two-level quantum system with a PT-symmetric non-Hermitian Hamiltonian, we have explicitly calculated the transition amplitudes ${\cal A}^{\cal CPT}_{\alpha \beta} \equiv \langle \widetilde{u}^{}_\beta|u^{}_\alpha(t)\rangle$ by introducing the ${\cal CPT}$ eigenstates $|\widetilde{u}^{}_\beta\rangle \equiv \left[|u^{}_\beta\rangle + {\cal CPT}|u^{}_\beta\rangle\right]/2$, where the flavor indices $\{\alpha, \beta\}$ are now running over $\{a, b\}$ and ${\cal CPT}$ stands for the combined charge-conjugate, parity, and time-reversal operator. The properly normalized transition probabilities ${\cal P}^{\cal CPT}_{\alpha \beta}$ associated with those amplitudes ${\cal A}^{\cal CPT}_{\alpha \beta}$ are then demonstrated to be conserved in the sense that ${\cal P}^{\cal CPT}_{aa} + {\cal P}^{\cal CPT}_{ab} = 1$ and ${\cal P}^{\cal CPT}_{ba} + {\cal P}^{\cal CPT}_{bb} = 1$. Although the explicit calculations are performed only for the two-level quantum system, it is straightforward to extend them to more general cases.

Finally, this formalism has been applied to two-flavor neutrino oscillations in vacuum, namely, $|\nu^{}_\mu\rangle \to |\widetilde{\nu}^{}_\mu\rangle$ and $|\nu^{}_\mu\rangle \to |\widetilde{\nu}^{}_\tau\rangle$. The Hamiltonian for two-flavor neutrino oscillations in vacuum becomes PT-symmetric only if the neutrino mixing angle is maximal, i.e., $\theta^{}_{23} = \pi/4$, which is compatible with current data of neutrino oscillation experiments. Therefore, the most general PT-symmetric non-Hermitian Hamiltonian for two-flavor neutrino oscillations in vacuum can be written as the ordinary Hermitian Hamiltonian with a maximal mixing angle plus PT-symmetric non-Hermitian corrections. It turns out that non-Hermitian parameters contribute a constant shift in the oscillation phase and even change the oscillation frequency, which could be tightly constrained by future precision measurements of neutrino oscillation parameters. Furthermore, in ordinary two-flavor neutrino physics, neutrino oscillations arise if and only if neutrinos have mass. However, in neutrino model building, using a PT-symmetric non-Hermitian Hamiltonian for two neutrino flavors, neutrino oscillations could arise, even though no mass for neutrinos have been introduced, i.e., there are no parameters in the Hamiltonian that describe neutrino mass in the conventional sense. Therefore, from a neutrino model-building point of view, it is of great importance to investigate two-level quantum systems with PT-symmetric non-Hermitian Hamiltonians and to find the transition probabilities between the two levels. As far as three-flavor neutrino oscillations are concerned, it is interesting to observe that the PT symmetry may be identified with the so-called $\mu$-$\tau$ reflection symmetry in the neutrino sector~\cite{Xing:2015fdg}, which is phenomenologically favored at present.

We believe that the discussion on conservation of transition probabilities and the explicit calculations in the present work could help to understand the main features of PT-symmetric non-Hermitian Hamiltonians. It remains to be observed if the newly introduced ${\cal CPT}$ eigenstate can be realized in some of the physical systems, where PT-symmetric non-Hermitian Hamiltonians have found intriguing applications.

\section*{Acknowledgments}
T.O.~acknowledges support by the Swedish Research Council (Vetenskapsr{\aa}det) through contract No.~2017-03934 and the KTH Royal Institute of Technology for a sabbatical period at the Uni\-versity of Iceland. The work of S.Z.~was supported in part by the National Natural Science Foundation of China under grant No.~11775232 and No.~11835013, and by the CAS Center for Excellence in Particle Physics.

\vspace{0.5cm}

\appendix

\section{Hermitian Quantum Mechanics}\label{sec:AppA}

In ordinary quantum mechanics, the wave functions or the state vectors $|\psi\rangle$ (in Dirac's bra-ket notations) are used to describe the state of a quantum system. The complex vector space, which is spanned by the state vectors, can be further endowed with a proper inner product, forming the Hilbert space $\mathfrak{H}$. The time evolution of the state vectors is governed by the Hamiltonian operator ${\cal H}$, which is usually assumed to be Hermitian with respect to the inner product defined as
\begin{equation}\label{eq:inner1}
\langle \psi|\phi\rangle \equiv |\psi\rangle^\dagger \cdot |\phi\rangle \; ,
\end{equation}
where $\langle \psi| \equiv |\psi\rangle^\dagger$ with the dagger ``$\dagger$" being complex conjugate and matrix transpose, and ``$\cdot$" denotes ordinary matrix multiplication. The exact meaning of matrix transpose and multiplication will become clear later.

For clarity, we only consider a finite-dimensional Hilbert space, for which one can always find a linearly independent and complete set of state vectors $|e^{}_i\rangle$ (for $i = 1, 2, \cdots, N$ with $N$ being a positive integer) and any state vector $|\psi\rangle$ can then be expressed as a linear superposition of them, viz.,
\begin{equation}
|\psi\rangle = \sum^N_{i=1} c^{}_i |e^{}_i\rangle \; ,
\end{equation}
where $c^{}_i$ (for $i = 1, 2, \cdots, N$) are complex coefficients. It is convenient to choose the set of basis vectors $\{|e^{}_i\rangle\}_{i=1}^N$ to be an orthonormal basis such that the column vectors $|e^{}_i\rangle$ (for $i = 1, 2, \cdots, N$) are orthonormal to each other, i.e., $\left[|e^{}_i\rangle\right]^{}_j = \delta^{}_{ij}$, where $\left[|e^{}_i\rangle\right]^{}_j$ is the $j$-th component of the $i$-th column vector $|e^{}_i\rangle$ and $\delta^{}_{ij}$ is the Kronecker delta. In this orthonormal basis, we have $\langle e^{}_i|e^{}_j\rangle = \delta^{}_{ij}$, according to the Euclidean inner product given in Eq.~(\ref{eq:inner1}). Any operator ${\cal O}$ in the Hilbert space $\mathfrak{H}$ is acting on the state vectors and can be represented by a matrix $O$, whose matrix elements are given by
\begin{equation}\label{eq:Omatrix}
O^{}_{mn} \equiv \langle e^{}_m|{\cal O}|e^{}_n \rangle \; ,
\end{equation}
for $m, n = 1, 2, \cdots, N$. The Hamiltonian operator ${\cal H}$ is supposed to be Hermitian, and thus, the corresponding representation matrix $H$ is Hermitian, i.e.,
\begin{equation}
H^{}_{mn} \equiv \langle e^{}_m|{\cal H}|e^{}_n\rangle = \langle e^{}_n|{\cal H}|e^{}_m\rangle^* = H^*_{nm} \; .
\end{equation}
The Hermiticity of the operators, including the Hamiltonian ${\cal H}$, should be understood in a more general way, namely, $\langle \psi|{\cal O}|\phi\rangle = \langle \phi|{\cal O}|\psi\rangle^*$, for any two arbitrary state vectors $|\phi\rangle$ and $|\psi\rangle$, or equivalently, we have $\langle {\cal O}\psi|\phi\rangle = \langle \psi|{\cal O}\phi\rangle$ or ${\cal O}^\dagger = {\cal O}$, which is called self-adjoint or Hermitian with respect to the Euclidean inner product.

The Hermitian Hamiltonian matrix $H$ can be diagonalized via the similarity transformation $U H U^{-1} = \widehat{H}$ with a unitary matrix $U$, i.e., $U^{-1} = U^\dagger$, where $\widehat{H} = {\rm diag}(E^{}_1, E^{}_2, \cdots, E^{}_N)$ is a diagonal matrix with $E^{}_i$ (for $i=1, 2, \cdots, N$) being real energy eigenvalues. If the corresponding energy eigenvectors are denoted by $|\epsilon^{}_i\rangle$, namely, $H|\epsilon^{}_i\rangle = E^{}_i |\epsilon^{}_i\rangle$, then $|\epsilon^{}_i\rangle$ can be identified as the $i$-th column of the unitary matrix $U^{-1}$. Obviously, the orthonormality conditions $\langle \epsilon^{}_i|\epsilon^{}_j\rangle = \delta^{}_{ij}$ (for $i, j = 1, 2, \cdots, N$) are satisfied due to the unitarity of $U$, namely, $U^\dagger U = \mathbb{1}_N$. Some remarks on the basis transformation and the time evolution of the state vectors are helpful.
\begin{itemize}
\item Since $H$ and $\widehat{H}$ are the representation matrices of ${\cal H}$ in the bases $\{|e^{}_i\rangle \}^{N}_{i=1}$ and $\{|\epsilon^{}_i\rangle\}^{N}_{i=1}$, respectively, it is then straightforward to establish the connection between these two sets of state vectors. More explicitly, we have
    \begin{equation}
    \langle e^{}_i|{\cal H}|e^{}_j\rangle = H^{}_{ij} = \sum_{m=1}^N \sum_{n=1}^N (U^{-1})^{}_{i m} \widehat{H}^{}_{mn} U^{}_{nj} = \sum_{m=1}^N \sum_{n=1}^N (U^{-1})^{}_{im} \langle \epsilon^{}_m| {\cal H} |\epsilon^{}_n\rangle U^{}_{nj} \; ,
    \end{equation}
    and thus, $|e^{}_j\rangle = \sum^{N}_{n=1} (U^{\rm T})^{}_{jn} |\epsilon^{}_n\rangle$ or $|\epsilon^{}_n\rangle = \sum^{N}_{j=1} U^*_{nj}|e^{}_j\rangle$. On the other hand, one can also recast such a basis transformation into matrix form
    \begin{equation}\label{eq:transform}
    (|e^{}_1\rangle, |e^{}_2\rangle, \cdots, |e^{}_N\rangle) = (|\epsilon^{}_1\rangle, |\epsilon^{}_2\rangle, \cdots, |\epsilon^{}_N\rangle) \cdot U \; ,
    \end{equation}
    where $|e^{}_i\rangle$ and $|\epsilon^{}_i\rangle$ (for $i = 1, 2, \cdots, N$) are all column vectors. It is easy to observe that it is just the identity matrix $(|e^{}_1\rangle, |e^{}_2\rangle, \cdots, |e^{}_N\rangle) = \mathbb{1}_N$ on the left-hand side of Eq.~(\ref{eq:transform}), which holds trivially because of $(|\epsilon^{}_1\rangle, |\epsilon^{}_2\rangle, \cdots, |\epsilon^{}_N\rangle) = U^{-1}$ by construction. Adopting the basis $\{|e^{}_i\rangle\}_{i=1}^N$, one can find the representation matrices for all operators and state vectors. Therefore, any similarity transformation of $O$ via $O^\prime = V O V^{-1}$, where $V$ is a unitary matrix, can be compensated by the transformation of the state vectors $|\psi\rangle$ through $|\psi^\prime\rangle = V |\psi\rangle$ such that the expectation value $\langle \psi|O|\psi\rangle = \langle \psi^\prime| O^\prime |\psi^\prime\rangle$ is unchanged.

\item For later convenience, we ignore the dependence of the state vectors on the spatial coordinates and assume that the Hamiltonian is also time-independent. In this case, the time evolution of the state vector $|\psi(t)\rangle$ is governed by the Schr\"{o}dinger equation
\begin{equation}\label{eq:Schroedinger}
{\rm i}\frac{{\rm d}}{{\rm d}t} |\psi(t)\rangle = {\cal H} |\psi(t)\rangle \; ,
\end{equation}
which can be rewritten the matrix form as
\begin{equation}\label{eq:mSchroe}
{\rm i} \frac{{\rm d}}{{\rm d}t} \left(\begin{matrix} c^{}_1(t) \cr c^{}_2(t) \cr \vdots \cr c^{}_N(t) \end{matrix}\right) = H \left(\begin{matrix} c^{}_1(t) \cr c^{}_2(t) \cr \vdots \cr c^{}_N(t) \end{matrix}\right) \; ,
\end{equation}
where $|\psi(t)\rangle \equiv \sum^N_{i=1} c^{}_i(t) |e^{}_i\rangle$ with $c^{}_i(t)$ being complex functions of time $t$. Now, let us look at the time evolution of the inner product of $|\psi(t)\rangle$ and another arbitrary state vector $|\phi(t)\rangle = \sum_{i=1}^N c^\prime_i(t) |e^{}_i\rangle$, namely, $\langle \psi(t)|\phi(t)\rangle = \sum_{i=1}^N c^*_i(t) c^\prime_i(t)$, which at time $t = 0$ is given by $\langle\psi(0)|\phi(0)\rangle = \sum_{i=1}^N c^*_i(0) c^\prime_i(0)$. According to Eq.~(\ref{eq:mSchroe}), we have
\begin{equation}\label{eq:norm1}
{\rm i}\frac{{\rm d}}{{\rm d}t} \langle \psi(t)|\phi(t)\rangle = ( c^*_1(t), c^*_2(t), \cdots, c^*_N(t) ) \cdot (H - H^\dagger) \cdot \left(\begin{matrix} c^\prime_1(t) \cr c^\prime_2(t) \cr \vdots \cr c^\prime_N(t) \end{matrix}\right) \; ,
\end{equation}
which becomes time-independent only if $H$ is Hermitian, i.e.,  $H^\dagger = H$. As usual, the transition amplitude for $|e^{}_i\rangle \to |e^{}_j\rangle$ refers to the probability that the system is initially prepared in the state $|e^{}_i(0)\rangle \equiv |e^{}_i\rangle$ at time $t = 0$, and then evolves to $|e^{}_i(t)\rangle$ at time $t$ with a probability to be in the state $|e^{}_j\rangle$. Following the time evolution of $|e^{}_i(t)\rangle$ according to the Schr\"{o}dinger equation, we can figure out the amplitude ${\cal A}^{}_{i j} \equiv \langle e^{}_j|e^{}_i(t)\rangle = \langle e^{}_j|e^{-{\rm i}H t}|e^{}_i\rangle$, and then, the probability ${\cal P}^{}_{ij} \equiv \left|{\cal A}^{}_{ij}\right|^2$. The unitarity of the theory can be checked by summing over the final-state index of the transition probabilities
\begin{equation}\label{eq:unitarity}
\sum_{j=1}^N {\cal P}^{}_{ij} = \sum_{j=1}^N \left|{\cal A}^{}_{ij}\right|^2 = \sum_{j=1}^N \langle e^{}_i(t)|e^{}_j\rangle \langle e^{}_j|e^{}_i(t)\rangle = \langle e^{}_i(t)|e^{}_i(t)\rangle = \langle e^{}_i|e^{{\rm i}(H^\dagger - H)t}|e^{}_i\rangle \;,
\end{equation}
which turns out to be unity when $H^\dagger = H$ is satisfied. Note that the completeness relation $\sum_{j=1} |e^{}_j\rangle \langle e^{}_j| = \mathbb{1}_N$ has been utilized in Eq.~(\ref{eq:unitarity}), where one can observe that it is the completeness relation and the Hermiticity of the Hamiltonian $H$ that together leads to conservation of probability or the unitarity of the whole theory.
\end{itemize}

\section{Non-Hermitian Quantum Mechanics}\label{sec:AppB}

All the above points are quite standard in ordinary quantum mechanics, but we think it is necessary and helpful to collect them such that one can make a close comparison with the results in non-Hermitian quantum mechanics. If we relax the Hermiticity of the Hamiltonian $H$, then additional requirements should be imposed to ensure a real spectrum of the system. It has been proven in Refs.~\cite{Mostafazadeh:2001jk, Mostafazadeh:2001nr, Mostafazadeh:2002id} that the necessary and sufficient conditions for a non-Hermitian but diagonalizable Hamiltonian to have real eigenvalues is the existence of a linear positive-definite operator $\eta^{}_+$ such that $\eta^{}_+ H \eta^{-1}_+ = H^\dagger$ is fulfilled. If this is the case, then $H$ is called $\eta^{}_+$-pseudo-Hermitian. In other words, we can construct another Hilbert space $\mathfrak{H}^*$ that shares the same state vectors as $\mathfrak{H}$, but is now endowed with a different inner product, viz.,
\begin{equation}\label{eq:inner2}
\langle \psi|\phi\rangle^{}_+ \equiv \langle \psi| \eta^{}_+ |\phi\rangle = |\psi\rangle^\dagger \cdot \eta^{}_+ \cdot |\phi\rangle \; ,
\end{equation}
which should be compared with the Euclidean inner product in Eq.~(\ref{eq:inner1}). In $\mathfrak{H}^*$ with this new inner product, one can immediately verify that
\begin{equation}\label{eq:pseudoH}
\langle H \psi|\phi\rangle^{}_+ = \left(H|\psi\rangle\right)^\dagger \cdot \eta^{}_+ \cdot |\phi\rangle = |\psi\rangle^\dagger \cdot H^\dagger \eta^{}_+ \cdot |\phi\rangle = \langle \psi|H \phi\rangle^{}_+ \; ,
\end{equation}
where the pseudo-Hermiticity relation $\eta^{}_+ H \eta^{-1}_+ = H^\dagger$ has been used. Comparing Eq.~(\ref{eq:pseudoH}) with the Hermiticity condition $\langle H \psi|\phi\rangle = \langle \psi|H\phi\rangle$ in ordinary quantum mechanics, we observe that $H$ with the Euclidean inner product is actually Hermitian with respect to the $\eta^{}_+$ inner product~\cite{Mostafazadeh:2004mx}. Some important comments on non-Hermitian quantum mechanics are in order.
\begin{itemize}
\item Since the Hamiltonian $H$ is supposed to be diagonalizable, we denote the eigenvalues of $H$ by $E^{}_i$ and the corresponding eigenvectors by $|\psi^{}_i\rangle$, i.e., $H |\psi^{}_i\rangle = E^{}_i |\psi^{}_i\rangle$ (for $i = 1, 2, \cdots, N$). Due to the relation $\eta^{}_+ H \eta^{-1}_+ = H^\dagger$, one can construct another set of state vectors $|\phi^{}_i\rangle \equiv \eta^{}_+ |\psi^{}_i\rangle$ such that $H^\dagger |\phi^{}_i\rangle = E^{}_i |\phi^{}_i\rangle$ (for $i = 1, 2, \cdots, N$). It is easy to prove the following orthonormality conditions and completeness relation
    \begin{equation}
    \langle \phi^{}_i |\psi^{}_j\rangle = \delta^{}_{ij} \; , \qquad \sum_{i=1}^N |\psi^{}_i\rangle \langle \phi^{}_i| = \mathbb{1}_N \; .
    \end{equation}
    Using the above completeness relation and the definition $|\phi^{}_i\rangle \equiv \eta^{}_+|\psi^{}_i\rangle$, we can find the positive-definite metric operator $\eta^{}_+$ and its inverse as~\cite{Kleefeld:2009vd}
    \begin{equation}\label{eq:etapm}
    \eta^{}_+ = \sum_{i=1}^N |\phi^{}_i\rangle \langle \phi^{}_i| \; , \qquad \eta^{-1}_+ = \sum_{i=1}^N |\psi^{}_i\rangle \langle \psi^{}_i| \; .
    \end{equation}
    Obviously, these results are of practical use to construct $\eta^{}_+$ and $\eta^{-1}_+$.

\item Now, we are in the position to investigate the time evolution of the state vectors in the non-Hermitian framework. Similar to ordinary quantum mechanics, the state vector $|\psi(t)\rangle$ evolves in time according to the Schr\"{o}dinger equation. Consider the time derivative of the norm $\langle \psi(t)|\phi(t)\rangle^{}_+$ for two arbitrary state vectors
    \begin{equation}\label{eq:norm2}
    {\rm i}\frac{{\rm d}}{{\rm d}t} \langle \psi(t)|\phi(t)\rangle^{}_+ = \langle \psi(t)| \left(\eta^{}_+ H - H^\dagger \eta^{}_+\right) |\phi(t)\rangle \; ,
    \end{equation}
    which vanishes using the condition $\eta^{}_+ H \eta^{-1}_+ = H^\dagger$. Therefore, the norms of state vectors with respect to the $\eta^{}_+$ inner product are time-independent, which should be compared with the result in Eq.~(\ref{eq:norm1}) for ordinary quantum mechanics.

\item Finally, we examine the transition amplitudes ${\cal A}^{+}_{ij} \equiv \langle e^{}_j|e^{}_i(t)\rangle^{}_+$ and the relevant probabilities ${\cal P}^{+}_{ij} \equiv \left|{\cal A}^{+}_{ij}\right|^2$, and then check the unitarity of the non-Hermitian theory. Let us look at the summation of probabilities
    \begin{equation}\label{eq:prob2}
    \sum_{j=1}^N {\cal P}^{+}_{ij} = \sum_{j=1}^N \langle e^{}_i(t)|\eta^{}_+ |e^{}_j\rangle \langle e^{}_j|\eta^{}_+|e^{}_i(t)\rangle = \langle e^{}_i(t)|\eta^{}_+|e^{}_i(t)\rangle^{}_+ \; ,
    \end{equation}
    where $\sum_{j=1}^N |e^{}_j\rangle \langle e^{}_j| = \mathbb{1}_N$ has been used in the last step. The rightmost result in Eq.~(\ref{eq:prob2}) can also be written as $\langle e^{}_i(t)|\eta^{}_+|e^{}_i(t)\rangle^{}_+ = \langle e^{}_i(t)|\eta^2_+|e^{}_i(t)\rangle$, whose time derivative can be found below
    \begin{equation}\label{eq:probtime}
    {\rm i}\frac{{\rm d}}{{\rm d}t} \langle e^{}_i(t)|\eta^2_+|e^{}_i(t)\rangle = \langle e^{}_i(t)|\left(\eta^2_+ H - H^\dagger \eta^2_+\right)|e^{}_i(t)\rangle = \langle e^{}_i(t)|\left\{\eta^{}_+ \left[\eta^{}_+, H\right]\right\}|e^{}_i(t)\rangle\; ,
    \end{equation}
    where the commutator $[\eta^{}_+, H] \equiv \eta^{}_+ H - H \eta^{}_+$ has been defined. If we demand that the summation of transition probabilities over the final states should be time-independent, then the Hamiltonian $H$ possesses a symmetry represented by the metric operator $\eta^{}_+$, i.e., $[\eta^{}_+, H] = {\mathbb 0}_N$ or $\eta^{}_+ H \eta^{-1}_+ = H$. A comparison with $\eta^{}_+ H \eta^{-1}_+ = H^\dagger$ leads to the condition $H^\dagger = H$, namely, $H$ must be Hermitian. Therefore, the above definitions of transition amplitudes ${\cal A}^+_{ij}$ and probabilities ${\cal P}^{+}_{ij}$ should be revisited. See, e.g., Refs.~\cite{Bagarello1, Bagarello2}, for an earlier attempt in this connection.
\end{itemize}
So far, we have recalled the well-known results in Hermitian and pseudo-Hermitian quantum mechanics, establishing our notations and conventions for later discussion. However, the transition probabilities in the pseudo-Hermitian or PT-symmetric Hamiltonian framework have not yet been studied in the literature, which will be of our main concern in the rest of this work.

Finally, for the two-level quantum system discussed in Sec.~\ref{sec:2}, we collect other possible definitions of inner products for completeness.
\begin{itemize}
\item {\bf Orthogonal inner product.} --- This is defined by assuming $\eta = {\bf 1}$ in Eq.~(\ref{eq:eta}). In this case, it is interesting to show that
    \begin{equation}
    \left( \begin{matrix} \langle u^{}_+|u^{}_+\rangle^{}_{\bf 1} & \langle u^{}_+|u^{}_-\rangle^{}_{\bf 1} \cr \langle u^{}_-|u^{}_+\rangle^{}_{\bf 1} & \langle u^{}_-|u^{}_-\rangle^{}_{\bf 1}\end{matrix} \right) = \left(\begin{matrix} 1 & 0 \cr 0 & 1 \end{matrix}\right) \; ,
    \quad
     \left( \begin{matrix} \langle u^{}_a|u^{}_a\rangle^{}_{\bf 1} & \langle u^{}_a|u^{}_b\rangle^{}_{\bf 1} \cr \langle u^{}_b|u^{}_a\rangle^{}_{\bf 1} & \langle u^{}_b|u^{}_b\rangle^{}_{\bf 1}\end{matrix} \right) = \left(\begin{matrix} 1 & 0 \cr 0 & 1\end{matrix}\right) \; .
    \end{equation}

\item {\bf ${\cal P}$ inner product.} --- Using $\eta = {\cal P}$, one can verify that
    \begin{equation}
     \left( \begin{matrix} \langle u^{}_+|u^{}_+\rangle^{}_{\cal P} & \langle u^{}_+|u^{}_-\rangle^{}_{\cal P} \cr \langle u^{}_-|u^{}_+\rangle^{}_{\cal P} & \langle u^{}_-|u^{}_-\rangle^{}_{\cal P}\end{matrix} \right) = \left(\begin{matrix} \sec \alpha & -{\rm i}\tan\alpha \cr -{\rm i}\tan\alpha & -\sec\alpha \end{matrix}\right) \; ,
    \quad
    \left( \begin{matrix} \langle u^{}_a|u^{}_a\rangle^{}_{\cal P} & \langle u^{}_a|u^{}_b\rangle^{}_{\cal P} \cr \langle u^{}_b|u^{}_a\rangle^{}_{\cal P} & \langle u^{}_b|u^{}_b\rangle^{}_{\cal P}\end{matrix} \right) = \left(\begin{matrix} 0 & 1 \cr 1 & 0\end{matrix}\right) \; .
    \end{equation}

\item {\bf ${\cal PT}$ inner product.} --- Given $\eta = {\cal PT}$, one can see that for the real ``flavor'' eigenstate vectors $|u^{}_a\rangle$ and $|u^{}_b\rangle$, the ${\cal PT}$ inner product is equivalent to the ${\cal P}$ inner product. For the ``mass" eigenstate vectors $|u^{}_+\rangle$ and $|u^{}_-\rangle$, we have already chosen to normalize them by using the ${\cal PT}$ inner product. In this case, we have
     \begin{equation}\label{eq:PTnorm}
    \left( \begin{matrix} \langle u^{}_+|u^{}_+\rangle^{}_{\cal PT} & \langle u^{}_+|u^{}_-\rangle^{}_{\cal PT} \cr \langle u^{}_-|u^{}_+\rangle^{}_{\cal PT} & \langle u^{}_-|u^{}_-\rangle^{}_{\cal PT}\end{matrix} \right) = \left(\begin{matrix} 1 & 0 \cr 0 & -1 \end{matrix}\right) \; ,
    \qquad
    \left( \begin{matrix} \langle u^{}_a|u^{}_a\rangle^{}_{\cal PT} & \langle u^{}_a|u^{}_b\rangle^{}_{\cal PT} \cr \langle u^{}_b|u^{}_a\rangle^{}_{\cal PT} & \langle u^{}_b|u^{}_b\rangle^{}_{\cal PT}\end{matrix} \right) = \left(\begin{matrix} 0 & 1 \cr 1 & 0\end{matrix}\right) \; .
    \end{equation}
\end{itemize}

\end{document}